# Converting Alzheimer's disease map into a heavyweight ontology: a formal network to integrate data

Vincent Henry[1,2], Ivan Moszer[2], Olivier Dameron[3], Marie-Claude Potier[2], Martin Hofmann-Apitius[4] and Olivier Colliot[2,1,5]

[1] Inria, Aramis project-team, Paris, France
[2] ICM, Inserm U1127, CNRS UMR 7225, Sorbonne Université, Paris, France
[3] Univ Rennes, CNRS, Inria, IRISA - UMR 6074, F-35000 Rennes, France
[4] Fraunhofer SCAI, Sankt Augustin, Germany
[5] AP-HP, Pitié-Salpêtrière Hospital, Dep. of Neurology and Neuroradiology, Paris, France
vincent.henry@inria.fr – olivier.colliot@upmc.fr

**Abstract.** Alzheimer's disease (AD) pathophysiology is still imperfectly understood and current paradigms have not led to curative outcome. Omics technologies offer great promises for improving our understanding and generating new hypotheses. However, integration and interpretation of such data pose major challenges, calling for adequate knowledge models. AlzPathway is a disease map that gives a detailed and broad account of AD pathophysiology. However, AlzPathway lacks formalism, which can lead to ambiguity and misinterpretation. Ontologies are an adequate framework to overcome this limitation, through their axiomatic definitions and logical reasoning properties. We introduce the AD Map Ontology (ADMO) an ontological upper model based on systems biology terms. We then propose to convert AlzPathway into an ontology and to integrate it into ADMO. We demonstrate that it allows one to deal with issues related to redundancy, naming, consistency, process classification and pathway relationships. Further, it opens opportunities to expand the model using elements from other resources, such as generic pathways from Reactome or clinical features contained in the ADO (AD Ontology). A version of the ontology will be made freely available to the community on Bioportal at the time of the conference.

## 1 Introduction

Alzheimer's disease (AD) is a progressive neurodegenerative disorder of the brain that was first described in 1906. The intense activity of AD research constantly generates new data and knowledge on AD-specific molecular and cellular processes (a Medline search for "Alzheimer disease" results in over 135,000 articles, as of June 30, 2018). However, the complexity of AD pathophysiology is still imperfectly understood [1]. These 110 years of efforts have essentially resulted in one dominant paradigm to underline the causes of AD: the amyloid cascade [2]. Therapeutics targeting this pathway failed to lead to curative outcome for humans, strongly suggesting the need for alternative hypotheses about AD etiology.

Since the turn of the century, omics technologies lead to a more comprehensive characterization of biological systems and diseases. The production of omics data in AD research opens promising perspectives to identify alternatives to the amyloid cascade paradigm. The current challenge is thus to integrate these data in an appropriate way, in order to propose new hypotheses and models about AD pathophysiology.

Systems medicine disease maps (DM) provide curated and integrated knowledge on pathophysiology of disorders at the molecular and phenotypic levels, which is adapted to the diversity of omics measurements [3]–[5]. Based on a systemic approach, they describe all biological physical entities (i.e. gene, mRNA, protein, metabolite) in their different states (e.g. phosphorylated protein, molecular complex, degraded molecule) and the interactions between them [6]. Their relations are represented as biochemical reactions organized in pathways, which encode the transition between participants' states as processes. AlzPathway is a DM developed for AD [3]. It describes 1,347 biological physical entities,129 phenotypes, 1,070 biochemical reactions and 26 pathways.

The information contained in DM is stored in syntactic formats developed for systems biology: the Systems Biology Graphical Notation (SBGN) [7] and the Systems Biology Markup Language (SBML) [8]. While syntactic formats are able to index information, they are not expressive enough to define explicit relationships and formal descriptions, leading to possible ambiguities and misinterpretations. For AlzPathway, this defect results in the lack of: a) hierarchy and disjunction between species, b) formal definition of phenotypes, c) formal relationships between reactions and pathways, d) uniformity of entities' naming and e) consistency between reactions and their participants.

Compared to syntactic formats, semantic formats used in ontologies, such as the Resource Description Framework (RDF) and the Web Ontology Language (OWL), have higher expressiveness [9], were designed to support integration and are thus good candidates to overcome the previous limitations.

An ontology is an explicit specification of a set of concepts and their relationships represented in a knowledge graph in semantic format. Ontologies provide a formal naming and definition of the types (i.e. the classes), properties, and interrelationships between entities that exist for a particular domain. Moreover, knowledge and data managed by an ontology benefit from its logical semantics and axiomatic properties (e.g. subsumption, disjunction, cardinality), which supports automatic control of consistency, automated enrichment of knowledge properties and complex query abilities [10].

The Alzheimer's Disease Ontology (ADO) [11] is the first ontology specific to the AD domain. ADO organizes information describing clinical, experimental and molecular features in OWL format. However, the description of the biological systems of ADO is less specific than that of AlzPathway.

Considering that 1) semantic formats can embed syntactic information, 2) DM provide an integrative view adapted to omics data management and 3) an ontological model is appropriate to finely manage data, the conversion of AlzPathway into a formal ontology would bring several assets, including an efficient integration of biomedical data for AD research, interconnection with ADO and an increased satisfiability of the resources.

We propose the Alzheimer Disease Map Ontology (ADMO), an ontological upper model able to embed the AlzPathway DM. Section 2 is devoted to the description of the ADMO model. In Section 3, we describe a method to convert AlzPathway in OWL and how ADMO can manage the converted AlzPathway and automatically enhance its formalism. Section 4 presents elements of discussion and perspectives.

## 2 Ontological upper model: Alzheimer Disease Map Ontology

The initial definition of an ontological model aims to design a knowledge graph that will drive its content. In a formal ontology, the relationships are not only links between classes, but also constraints that are inherited by all their descendants (subclasses). Thus, the choices of axioms that support high level classes and their properties are key elements for the utility of the model.

The Systems Biology Ontology (SBO) [12] is a terminology that provides a set of classes commonly used to index information in SBML format. These classes conceptualize biological entities at an adequate level of genericity and accuracy that supports a wide coverage with few classes and enough discrimination. We selected a set of 51 SBO classes for reactions, pathways and molecules as a first resource of subclasses of processes and participants, respectively. The modified Edinburg Pathway Notation (mEPN) [13] is another syntactic format based on systems approach. Its components provide a refined set of reactions that complete the SBO class set. Following class selection from SBO and mEPN, we designed a class hierarchy between them. We systematically added disjointness constraints between the sibling subclasses of participants in order to ensure that process participants belong to only one set (e.g. a gene cannot be a protein and reciprocally), We did not apply the same rule to the processes' subclasses were not disjoined as a reaction may refer to different processes (e.g. a transfer is an addition and a removal).

Properties consistent with a systems approach (i.e. *part_of*, *component_of*, *component_process_of*, *has_participant*, *has_input*, *has_output*, *has_active_participant*, *derives_from* and their respective inverse properties) were imported from the upper-level Relation Ontology (RO) [14]. Then, we formally defined our set of classes with these properties and cardinalities to link processes and participants with description logic (e.g. a transcription has at least one gene as input and has at least one mRNA as output; a protein complex formation has at least two proteins as input and has at least one protein complex as output).

The design of the ADMO upper ontological model based on SBO, mEPN and RO resulted in 77 classes (30 processes' subclasses and 38 participants subclasses) formally defined by 123 logical axioms in description logic (Fig. 1). This model is based on a simple pattern as our knowledge graph involves only three types of properties: 1) the *is_a* (*subclass_of*) standard property, 2) the *has_part* standard property and its sub-properties *has_component* and *has_component_process* and 3) the *has_participant* property and its sub-properties *has_input*, *has_output* and *has_active_participant*.

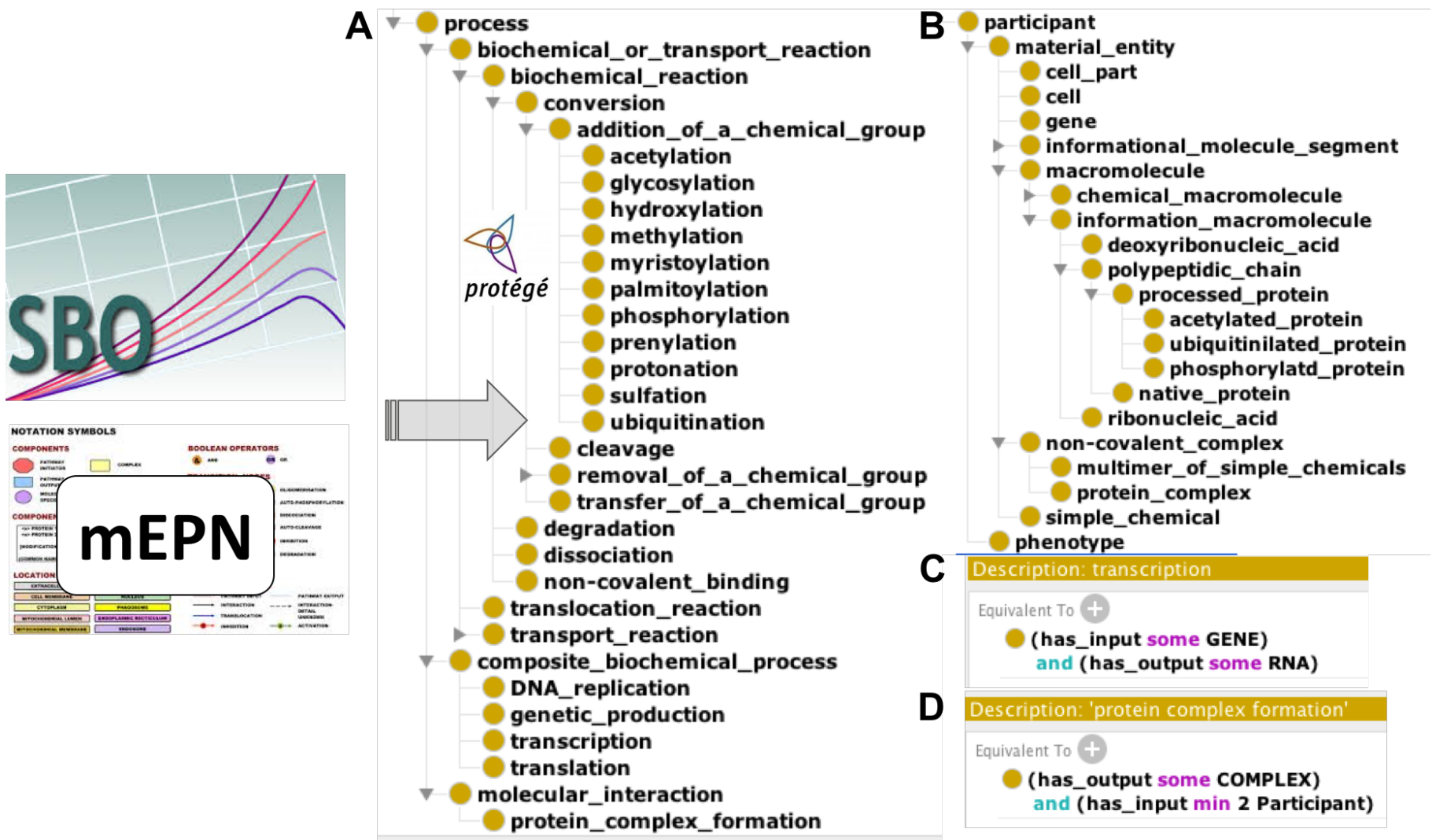

**Fig. 1. Alzheimer Disease Map Ontology model design.** Classes were extracted from the Systems Biology Ontology (SBO) and the modified Edinburg Pathway Notation (mEPN) into Protégé. Classes were hierarchized as subclasses of process (A) or participant (B). Using properties from the Relation Ontology (RO), classes were formally defined in description logic, as illustrated in the case of transcription (C) and protein complex formation (D) processes.

## 3 AlzPathway conversion and integration into ADMO

AlzPathway elements were extracted and stored in a structured table using home-made Python scripts. In this table, each biological entity was indexed by one of the high-level participants' subclasses of ADMO and all processes were in correspondence with their participants. The table also contains class annotations such as the AlzPathway identifier (ID), and IDs from other knowledge bases such as UniProt [15] for participants and KEGG [16] for processes. The table is structured to integrate component information for multiplex entities (e.g. protein complex) and location information for the process (e.g. cell type or cell part). The table was then manually curated as described below.

In AlzPathway, phenotypes are participants. But several of them are named with a process name, pathway label or molecule type (e.g. microglial activation, apoptosis or cytokines, respectively). In order to deal with these ambiguities, 26 phenotypes were reclassified as molecules (e.g. cytokine) or cellular component (e.g. membrane) and 14 names that referred to processes or pathways were changed into processes' participant names (e.g. apoptosis became apoptotic signal). In addition, 5 phenotypes that were named with a relevant pathway name (e.g. apoptosis) were added to the initial set of the 26 AlzPathway's pathways.

AlzPathway only describes a subset of genes, mRNA and proteins. As omics technology can capture data at the genome, transcriptome or proteome levels, we added missing information in order to complete some correspondences between genes and gene products. This resulted in the addition of 433 genes, 432 mRNA and 6 proteins.

Then, using the ontology editor Protégé, the content of the structured table was imported into ADMO using the Protégé Cellfie plugin. Entities information were integrated as subclasses of ADMO participants classes based on their naming in order to remove redundancies (120 of them were identified). During the integration, we also added a new property *has_template* (sub-property of *derives_from*) to formally link a gene to its related mRNA and a mRNA to its related protein. Reactions were integrated as independent subclasses. Then, automated reasoning was used to classify them as subclasses of the ADMO process classes depending on their formal definition (see Fig. 2a*). Inferred axioms were then edited. During their import, process classes from AlzPathway were formally linked to their respective location through the RO property: *occurs_in*.

While AlzPathway does not formally link pathways and their related biochemical reactions, pathways were manually imported. For each pathway, a class "reaction involved in pathway *x*" was created and both defined both as "reaction that *has_participant* the molecules of interest in *x*" and "*component_process_of* pathway *x*". For example, the class "reaction involved in WNT signaling pathway" *has_participant* "WNT" and is a *component_process_of* "WNT signaling pathway". Then, using automated reasoning, all reactions having participants involved in pathway *x* were classified as subclasses of "*component_process_of* pathway *x*" classes and were linked to the pathway with the *component_process_of* property by subsumption. For example, "SFRP-WNT association" is automatically classified as subclass of "reaction involved in WNT signaling pathway" (see Fig. 2b*) and inherits from its properties *component_process_of* "WNT signaling pathway" (see Fig. 2b**). Inferred axioms were then edited.

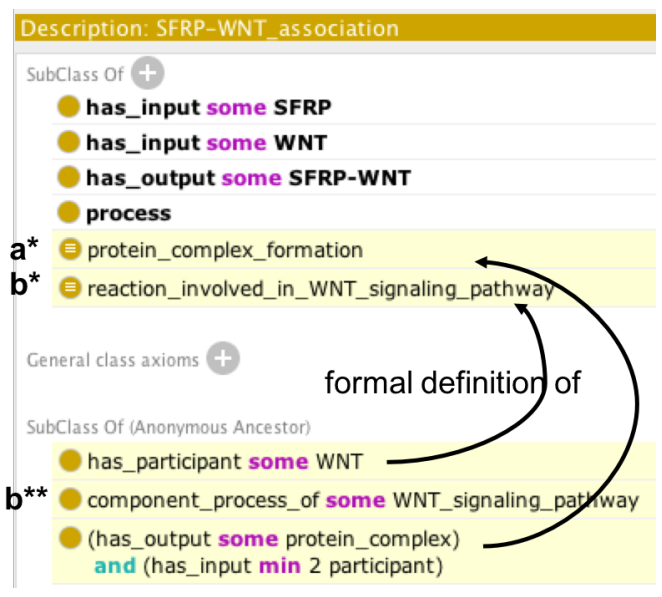

**Fig. 2. Example of automated reasoning on Protégé.** Asserted axioms are in uncoloured lines and inferred axioms are highlighted in yellow. Following automated reasoning SFRP-WNT heterodimer association is classified as subclass of protein complex formation (a*) and of reaction involved in WNT signaling pathway class (b*), thus it inherits of the *component_process_of* WNT_signaling pathway property (b**).

As a result, ADMO embeds AlzPathway in a consistent network containing 2132 classes (2128 participants, including 87 phenotypes and 1012 processes, including 46 pathways) in relation with 7814 logical axioms before and 8907 logical axioms after automated reasoning, respectively. Specific efforts were dedicated to the design of classes hierarchy and formal definition with description logic axioms, leading to explicit relations between biological entities and processes. These axioms were inherited by classes imported from AlzPathway, resulting in the formal and precise description of the elements of AD pathophysiology. The conversion of AlzPathway also benefits from the ADMO simple pattern of relationships (Fig. 3) in which new properties were added: the *derives_from* property that links a modified protein to its native form, the *has_template* property that links a native gene product to its mRNA and gene, and the *occurs_in* property that links a process to its cellular location. A version of the ontology will be made freely available to the community on Bioportal at the time of the conference.

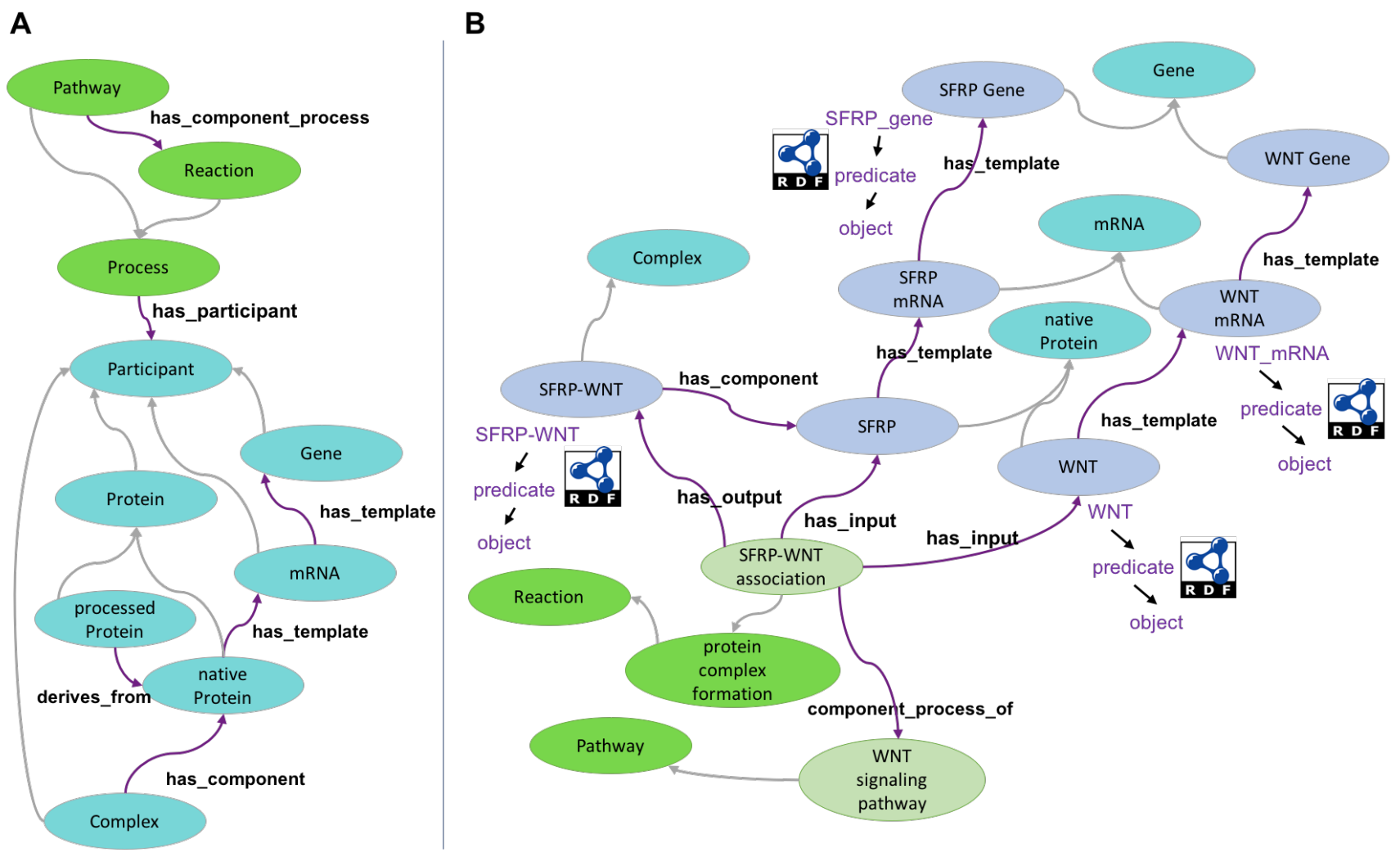

**Fig. 3. Alzheimer Disease Map Ontology (ADMO) pattern (A) and application to AlzPathway (B).** AlzPathway classes (B; illustrated for the SFRP-WNT association process ant its participants) are now subclasses of ADMO classes (A). Each class of AlzPathway may be instantiated by the corresponding entities as individuals. Then, entities can be related to different objects in an RDF schema such as patients and experiments, or more specifically to values such as SNP for genes, relative expression for mRNA, and concentration for proteins.

## 4 Discussion

We proposed the ADMO ontological model in order to manage the conversion and integration of AlzPathway in OWL format. By converting AlzPathway into an OWL ontology, we increased its formalism. All entities are now formally defined and interconnected within a consistent network. While AlzPathway contained several ambiguities, our efforts on formalism at a semantic level for phenotype and description logic in ADMO classes allowed us to solve inconsistencies. Moreover, the combination of SBO and mEPN provided a more precise specification of processes compared to SBML or SBGN, which was beneficial for the specification of AlzPathway reactions following its import into ADMO.

Unlike DM, ontologies are not adapted for graphical visualization but present a higher flexibility to integrate new elements in the knowledge graph, as we did by adding 865 genes and mRNA. Moreover, during the conversion step, AlzPathway's internal IDs were retained as class annotations, allowing interoperability between the initial and converted AlzPathway. Taking advantage of the knowledge graph and its semantic links, the ID information are retrievable from a derived molecule to its native form following the *derives_from* or *has_component* properties that link each of this classes.

Furthermore, the increased formalism requires to assert a participant as subclass of the most representative class and thus, clarifies the status of the entities. In several standard bioinformatics knowledge resources (e.g. UnitProt [15], KEGG[16]), a same ID refers to a gene or a protein and *in fine* to a set of information, such as gene, interaction, regulation and post translation modification (PTM), which are thus not specifically discriminated. However, current omics technologies are able to generate data focused on specific elements of the systems (gene mutation, relative gene expression, protein concentration…). This is underexploited by standard resources. Based on DM approaches, we provided an ontology that a) represents the complexity of a system such as AD pathophysiology and b) is designed to specifically integrate each type of omics data as an instance of the explicit corresponding class.

The next possible step is to instantiate the model with biomedical omics data. To this end, the RDF format is appropriate as it was specifically designed for representing a knowledge graph as a set of triples containing directed edges (semantic predicates). Different RDF schemas were already developed in the field of molecular biology (BioPax [17]) or more specifically for AD biomedical research (neuroRDF [18]). The Global Data Sharing in Alzheimer Disease Research initiative [19] is also a relevant resource to help find appropriate predicates to enrich RDF schemas and refine subject information (age, gender, clinical visit…). Depending on the need of a given study, users may design RDF schemas with their own predicates of interest. Then, this RDF schema can be integrated in our ontology by adding data as instances of its corresponding specific classes (Fig. 3B). Therefore, instantiation opens perspectives for complex querying; both richer and more precise than indexing.

DM are based on systems biology approaches, allowing one to take each part of the system into consideration. Our ontology goes one step further by formally defining the different elements of the system and linking them with the biochemical reaction and pathway levels. Here, we relied on AlzPathway, but additional resources could be used, such as Reactome [4] which provides a wide range of generic curated human biochemical reactions and pathways. Our ADMO upper ontological model provides an interesting framework to embed generic resources and thus harmonize AlzPathway and those resources. By converting and integrating AlzPathway in OWL format, the resulting ontology is ready to be connected with ADO and its clinical knowledge description. This offers new avenues for increasing the scale of representation of AD pathophysiology in our framework. In the same way, ADMO opens the perspective to harmonize specific DM from different neurodegenerative disorders such as the Parkinson's disease map [5] and others.

**Acknowledgements.** The research leading to these results has received funding from the program "Investissements d'avenir" ANR-10-IAIHU-06 (Agence Nationale de la Recherche-10-IA Institut Hospitalo-Universitaire-6) and from the Inria Project Lab Program (project Neuromarkers).